\documentclass[aps,pra,twocolumn]{revtex4}
\usepackage{graphicx,epsfig}
\usepackage{times}
\usepackage{amsmath,amssymb,wasysym}
\usepackage{color}
\usepackage[linktocpage=true,colorlinks=true, pdfborder={0 0 0},linkcolor=blue,citecolor=blue,urlcolor=blue,anchorcolor=black,linkcolor=blue]{hyperref}
\DeclareMathOperator{\rank}{rank}

\newcommand{\op}[1]{\hat{#1}}

\newcommand{\C}{\mathbb{C}}

\newcommand{\R}{\mathbb{R}}

\newcommand{\mean}[1]{\langle #1 \rangle}
\newcommand{\quoting}[1]{``#1''}
\newcommand{\vc}[1]{\vec{#1}}
\newcommand{\rem}[1]{}

\newcommand{\imag}[1]{\text{Im}(#1)}

\newcommand{\imagc}[1]{\text{Im}\,#1}

\newcommand{\real}[1]{\text{Re}(#1)}

\newcommand{\realc}[1]{\text{Re}\,#1}

\newcommand{\oHam}{\op{H}}
\newcommand{\oHami}{\op{\Gamma}}
\newcommand{\Hilbert}{{\cal H}}

\newcommand{\bra}[1]{\langle#1|}
\newcommand{\ket}[1]{|#1\rangle}

\newcommand{\Hform}[2]{(#1,#2)}
\newcommand{\braket}[2]{\langle#1|#2\rangle}
\newcommand{\braketff}[2]{\langle#1|#2\rangle_{\text{ff}}}

\newcommand{\sandwich}[3]{\langle#1|#2|#3\rangle}

\newcommand{\dHS}{d_{\text{HS}}}
\newcommand{\dph}{d_{\text{ph}}}
\newcommand{\dP}{d_{\text{P}}}
\newcommand{\vol}{\Omega}
\newcommand{\surf}{\partial\Omega}
\newcommand{\dsurf}{dA}
\newcommand{\ffA}{\zeta}
\newcommand{\ffB}{\xi}
\newcommand{\ffAv}{\vc{\zeta}}
\newcommand{\ffBv}{\vc{\xi}}
\newcommand{\normalv}{\vc{n}}
\newcommand{\couplingm}{\op{W}}
\newcommand{\ma}{l}
\newcommand{\mb}{j}
\newcommand{\decayrate}{\gamma}
\newcommand{\polar}{\theta}
\newcommand{\azimuthal}{\varphi}

\begin{document}

\title{Nonorthogonality constraints in open quantum and wave systems}
\author{Jan Wiersig}
\affiliation{Institut f{\"u}r Physik, Otto-von-Guericke-Universit{\"a}t Magdeburg, Postfach 4120, D-39016 Magdeburg, Germany}
\email{jan.wiersig@ovgu.de}
\date{\today}
\begin{abstract}
It is known that the squared modulus of the overlap (scalar product) of two energy eigenstates in a decaying quantum system is bounded from above by a function of the energy detuning and the individual decay rates. This is usually traced back to the positive definiteness of an appropriately defined decay operator. 
Here, we show that the weaker and more realistic condition of positive semi-definiteness is sufficient. We prove also that the bound becomes an equality for the case of single-channel decay. 
However, we show that the condition of positive semi-definiteness can be spoiled by quantum backflow. Hence, the overlap of quasibound quantum states subjected to outgoing-wave conditions can be larger than expected from the bound. A modified and less stringent bound, however, can be introduced. For electromagnetic systems, it turns out that a modification of the bound is not necessary due to the linear free-space dispersion relation. 
Finally, a geometric interpretation of the nonorthogonality bound is given which reveals that in this context the complex energy space can be seen as a surface of constant negative curvature.
\end{abstract}
\maketitle

\section{Introduction}
\label{sec:intro}
Any realistic quantum/wave system is an open system as it can never be perfectly isolated from its environment. The current strong interest in open systems arises from the new fields of non-Hermitian physics and parity-time (PT) symmetry~\cite{BB98,CW15,EMK18}. 
In this context open systems are often described by a non-Hermitian effective Hamiltonian~$\oHam$~\cite{Feshbach58,Feshbach62}, e.g., microwave cavities~\cite{SPK02}, ultracold atoms in optical lattices~\cite{KOA97,BerryODell98}, low-dimensional nanostructures~\cite{CK09}, PT-symmetric electronics~\cite{SLL12}, nuclear physics~\cite{MRW10}, optical microcavities~\cite{Wiersig11,KYW18}, and coupled cavity arrays~\cite{DM19}.
The non-Hermiticity or non-self-adjointness $\oHam \neq \oHam^\dagger$ (see Ref.~\cite{LPC18} for a rare case where a distinction of these two concepts is relevant in physics) implies that its eigenvalues~$E_\mb$ are complex-valued with the imaginary part determining a decay rate. If additionally the Hamiltonian is nonnormal, $[\oHam,\oHam^\dagger] \neq 0$, then its eigenstates are in general mutually nonorthogonal. The nonorthogonality is extreme near non-Hermitian degeneracies in parameter space, so-called exceptional points ~\cite{Kato66,Heiss00,Berry04,Heiss04,MA19}, where at least two eigenstates become collinear. At such points it is not possible to expand all states in terms of the eigenstates of $\oHam$. 

It is a common point of view that the nonorthogonality is just an inconvenience that can be brushed aside by using a proper biorthogonal basis. For time- (T) invariant systems this requires to use a bilinear form instead of the conventional scalar product~\cite{Moiseyev11}. An alternative point of view is that nonorthogonality signals or even represents interesting physics. For instance, nonorthogonality has been used to test the fundamental charge-parity-time (CPT)- and T-invariance of K$^0$ mesons~\cite{LOY57,LW65,BS66,BNS67}. Moreover, it has been shown that nonorthogonality can lead to nonexponential transient decay~\cite{TE05,MGT14,Wiersig18a}, chirality in perturbed whispering-gallery microcavities~\cite{WES11,Wiersig18b}, power oscillations in optical waveguides~\cite{MEC08,RMG10}, sensitivity of resonance widths under perturbation~\cite{FS12}, limitation of mode selectivity~\cite{DG18}, interesting transport properties~\cite{SWF04,DG19}, and quantum excess noise in lasers~\cite{Petermann79,Siegman86,SFP00,Schomerus09}.

It has been known for a long time that in decaying quantum systems there is an upper bound for the squared modulus of the overlap of two normalized energy eigenstates $\ket{\psi_\ma}$ and $\ket{\psi_\mb}$ 
\begin{equation}\label{eq:LWinequality}
|\braket{\psi_\ma}{\psi_\mb}|^2 \leq  \frac{\decayrate_\ma\decayrate_\mb}{\Delta_{\ma\mb}^2+(\decayrate_\ma+\decayrate_\mb)^2/4} \  ,
\end{equation}
with the conventional scalar product $\braket{\cdot}{\cdot}$, the energy detuning $\Delta_{\ma\mb} = \realc{E_\ma}-\realc{E_\mb}$, and the individual decay rates~$\decayrate_\mb = -2\,\imagc{E_\mb} \geq 0$. In particular, the inequality restricts the nonorthogonality to pairs of states which are \quoting{spectrally close}. 
Inequality~(\ref{eq:LWinequality}) was first published by Lee, Oehme, and Yang in 1957~\cite{LOY57} and later used among others by Lee and Wolfenstein~\cite{LW65}. It is therefore sometimes called the Lee-Wolfenstein inequality or bound. Often it is considered as part of the well-known Bell-Steinberger relations~\cite{BS66}, which is an equation for the overlap where the right-hand side is system-dependent but is bounded by the right-hand side of inequality~(\ref{eq:LWinequality}). 

Non-Hermitian Hamiltonians are commonly introduced in a phenomenological manner. Rarely, such an Hamiltonian is introduced from first principles, e.g., in Ref.~\cite{KYW18}. Instead of using an effective non-Hermitian Hamiltonian one frequently has to solve a wave equation with absorption/gain and outgoing-wave conditions. The solutions of such an equation correspond to the energy eigenstates and are called quasibound states, (quasi-normal) modes, or resonant states~\cite{Gamow28,KP38}. Inequality~(\ref{eq:LWinequality}) has not been proven directly for this case. 
In particular, it is not a priori clear how the normalization can be done as the quasibound states diverge in the far field, see, e.g.,~\cite{CD75}. There are many attempts to tackle the problem of normalization by using biorthogonality in terms of bilinear forms including surface integration terms instead of conventional scalar products based on volume integration, see, e.g.,~\cite{LLY94,CLM98,SHM13,KGH15,Perrin16}. 

The present paper adds a number of interesting and important results related to the Lee-Wolfenstein inequality. We clarify the requirements and demonstrate that for a certain kind of system the nonorthogonality becomes maximum in the sense that the inequality turns into an equality. Moreover, we show that the Lee-Wolfenstein inequality can fail for quasibound states based on wave equations with outgoing-wave conditions. Finally, we interpret the nonorthogonality constraint geometrically as an inequality between distances in Hilbert space and complex energy space.

The paper is organized as follows. Sections~\ref{sec:Heff} presents a derivation of the Lee-Wolfenstein inequality for systems described by an effective Hamiltonian. It is revealed for which systems the inequality becomes an equality. Moreover, we show how the inequality can be reformulated for biorthogonal basis sets. Section~\ref{sec:wave} deals with wave systems with absorption and outgoing-wave conditions, including quantum and electromagnetic systems. Section~\ref{sec:metric} provides a geometric interpretation. A discussion is presented in Sec.~\ref{sec:discussion}.

\section{Systems with effective Hamiltonian}
\label{sec:Heff}
\subsection{Derivation of the Lee-Wolfenstein inequality}
In this section we present an elegant derivation of the inequality~(\ref{eq:LWinequality}) for open systems which are described by an effective non-Hermitian Hamiltonian~$\oHam$. This derivation is inspired by Ref.~\cite{Urbanowski04} but it more explicitly and more correctly addresses the assumptions about~$\oHam$. 

We consider an $N$-dimensional Hilbert space~$\Hilbert$. The dynamics of any state vector $\ket{\psi}\in\Hilbert$ is determined by the Schr{\"o}dinger equation 
\begin{equation}\label{eq:SG}
i\hbar\frac{d}{dt}\ket{\psi} = \oHam\ket{\psi}
\end{equation}
with the reduced Planck's constant $\hbar$. Starting from the eigenproblem of~$\oHam$
\begin{equation}\label{eq:eigenproblem}
\oHam\ket{\psi_\mb} = E_\mb\ket{\psi_\mb}
\end{equation}
with the eigenstates $\ket{\psi_\mb}$ and the complex eigenvalues $E_\mb$ it is straightforward to show that
\begin{equation}\label{eq:e0}
-i(E_\ma^*-E_\mb)\braket{\psi_\ma}{\psi_\mb} = \sandwich{\psi_\ma}{\oHami}{\psi_\mb} \ ,
\end{equation}
where the raised asterisk denotes complex conjugation and $\oHami$ is the Hermitian decay matrix
\begin{equation}
\oHami = i(\oHam-\oHam^\dagger) \ .
\end{equation}
The mapping
\begin{equation}\label{eq:Hf}
(\phi,\psi) = \sandwich{\phi}{\oHami}{\psi} 
\end{equation}
is a so-called Hermitian form (or symmetric sesquilinear form)~\cite{Gallier11} because (i) it is linear in one argument (here the second, according to the standard physics notation for the scalar product) and (ii) symmetric under exchange of arguments and complex conjugation, i.e., $(\phi,\psi) = (\psi,\phi)^*$. 

The assumption needed to prove the Lee-Wolfenstein inequality is that the total probability $\braket{\psi}{\psi}$ decays monotonously, i.e.,
\begin{equation}\label{eq:dynamics}
\hbar\frac{d}{dt}\braket{\psi}{\psi} = -\sandwich{\psi}{\oHami}{\psi} = -(\psi,\psi) \leq 0
\end{equation}
for all $\ket{\psi} \in \Hilbert$. Hence, $\oHami$ is a {\em positive semi-definite} operator and $(\cdot,\cdot)$ is a {\em positive semi-definite} Hermitian form.
The stronger condition of strictly monotonously decaying probability would imply that $(\psi,\psi) > 0$ for all $\ket{\psi} \in \Hilbert$. In this case, $(\cdot,\cdot)$ is a positive-definite Hermitian form or, in other words, a scalar product. However, positive-definiteness as it is used for $\oHami$ in Refs.~\cite{LOY57,Urbanowski04} is not justified in general because a decaying quantum system can have non-decaying subspaces.

For a positive semi-definite Hermitian form, the Cauchy-Schwarz inequality holds, see, e.g.,~\cite{Gallier11},
\begin{equation}\label{eq:Cauchy}
|(\phi,\psi)|^2 \leq (\phi,\phi) (\psi,\psi)
\end{equation}
for all $\ket{\phi}, \ket{\psi} \in \Hilbert$. Applying the Cauchy-Schwarz inequality~(\ref{eq:Cauchy}) to Eq.~(\ref{eq:e0}) using Eq.~(\ref{eq:Hf}) we find
\begin{equation}\label{eq:e2}
\left[\Delta_{\ma\mb}^2+(\decayrate_\ma+\decayrate_\mb)^2/4\right]|\braket{\psi_\ma}{\psi_\mb}|^2 \leq (\psi_\ma,\psi_\ma) (\psi_\mb,\psi_\mb) 
\end{equation}
with the detuning $\Delta_{\ma\mb} = \realc{E_\ma}- \realc{E_\mb}$ and the decay rates $\decayrate_\mb = -2\,\imagc{E_\mb}$. The latter are nonnegative numbers as follows from Eqs.~(\ref{eq:e0}) and (\ref{eq:Hf}) 
\begin{equation}\label{eq:e3}
(\psi_\mb,\psi_\mb) = \decayrate_\mb 
\end{equation}
where we have used the normalization $\braket{\psi_\mb}{\psi_\mb} = 1$. From Eqs.~(\ref{eq:e2})-(\ref{eq:e3}) directly follow the Lee-Wolfenstein inequality~(\ref{eq:LWinequality}).

The basic ingredient for the derivation is the positive semi-definiteness of the Hermitian form $(\cdot,\cdot)$. This requirement is stronger than requiring only $\decayrate_\mb \geq 0$. In Ref.~\cite{Wiersig16} it was discussed that certain classes of decaying systems may only fulfill the latter condition ($\decayrate_\mb \geq 0$). For such kind of systems the inequality~(\ref{eq:LWinequality}) does not hold. We come back to this point in greater detail in Sec.~\ref{sec:wave}. 

It is an easy exercise for the reader to verify that the Lee-Wolfenstein inequality also holds for a decay matrix that is negative semi-definite. This applies to a monotonically increasing total probability in systems with gain but without loss. However, the Lee-Wolfenstein inequality is in general not valid for systems with gain {\em and} loss. An example may be a PT-symmetric system with balanced loss and gain at an exceptional point. Here, the two coalescing energy eigenstates $\ket{\psi_\ma}$ and $\ket{\psi_\mb}$ give $|\braket{\psi_\ma}{\psi_\mb}|^2 \to 1$ and $\decayrate_\ma = 0 = \decayrate_\mb$ which is clearly not consistent with the inequality~(\ref{eq:LWinequality}). 

\subsection{Dependence on the number of decay channels}
\label{sec:rank}
Are there systems for which the inequality~(\ref{eq:LWinequality}) is an equality for all energy eigenstates $\ket{\psi_\mb}$? According to the derivation in the previous subsection this question can be cast as the question of whether
\begin{equation}\label{eq:LWeq}
|(\psi_\ma,\psi_\mb)|^2 = (\psi_\ma,\psi_\ma) (\psi_\mb,\psi_\mb)
\end{equation}
is valid for all energy eigenstates. Equation~(\ref{eq:LWeq}) holds trivially if $\ket{\psi_\mb}$ is in the kernel of the decay matrix $\oHami$. In the complementary situation when $\ket{\psi_\mb}$ is in the span of $\oHami$ then Eq.~(\ref{eq:LWeq}) holds in general only if the span of $\oHami$ is one-dimensional. We conclude that the inequality~(\ref{eq:LWinequality}) turns into an equality when the rank of $\oHami$, which can be interpreted as the number of available decay channels, is unity. In this case the nonorthogonality is maximal for given detuning and decay rates. Examples of single-channel systems are tight-binding chains with a single lossy site~\cite{Wiersig18a} and doorway states in nuclear physics~\cite{SRS97}.
In the context of random-matrix theory it was already noticed by Fyodorov and Mehlig~\cite{FM02} that in the single-channel case the overlap of two energy eigenstates depends only on the complex eigenenergies, but the connection to the Lee-Wolfenstein inequality had not been established.

To obtain insight into the case of $\rank{\oHami} \neq 1$ we rewrite the Lee-Wolfenstein inequality~(\ref{eq:LWinequality}) as
\begin{equation}
\xi_{\ma\mb} \leq 1
\end{equation}
for all $\ma$ and $\mb$ by introducing the normalized bound $\xi_{\ma\mb} \geq 0$ such that
\begin{equation}\label{eq:LWxi}
|\braket{\psi_\ma}{\psi_\mb}|^2 = \xi_{\ma\mb} \frac{\decayrate_\ma\decayrate_\mb}{\Delta_{\ma\mb}^2+(\decayrate_\ma+\decayrate_\mb)^2/4} \ .
\end{equation}
Next we consider the average
\begin{equation}\label{eq:LWxiav}
\mean{\xi} = \frac{2}{N(N-1)}\sum_{\ma>\mb} \xi_{\ma\mb}
\end{equation}
as a measure of the nonnormality of the given system. The maximum value of $1$ is achieved for $\rank{\oHami} = 1$. To get a rough estimate of what is happening for higher rank, we first remark that 
\begin{equation}\label{eq:LWxiHf}
\xi_{\ma\mb} = \frac{|(\psi_\ma,\psi_\mb)|^2}{(\psi_\ma,\psi_\ma)(\psi_\mb,\psi_\mb)} \ ,
\end{equation}
assuming $(\psi_\ma,\psi_\ma) \neq 0 \neq (\psi_\mb,\psi_\mb)$. The only contributions come from the components in the span of $\oHami$ where the Hermitian form $(\cdot,\cdot)$ is positive definite. Restricted to the span of the decay operator it therefore is a scalar product. Next we assume that the eigenstates behave in an ergodic manner in Hilbert space. It is well known that for scalar products and ergodically distributed states the right-hand side of Eq.~(\ref{eq:LWxiHf}) is in average $1/M$ where $M$ is the dimension of the vector space. Here, $M = \rank{\oHami}$ and therefore
\begin{equation}\label{eq:xivsrank}
\mean{\xi} \approx \frac{1}{\rank{\oHami}} \ .
\end{equation}

\begin{figure}[ht]
\includegraphics[width=0.9\columnwidth]{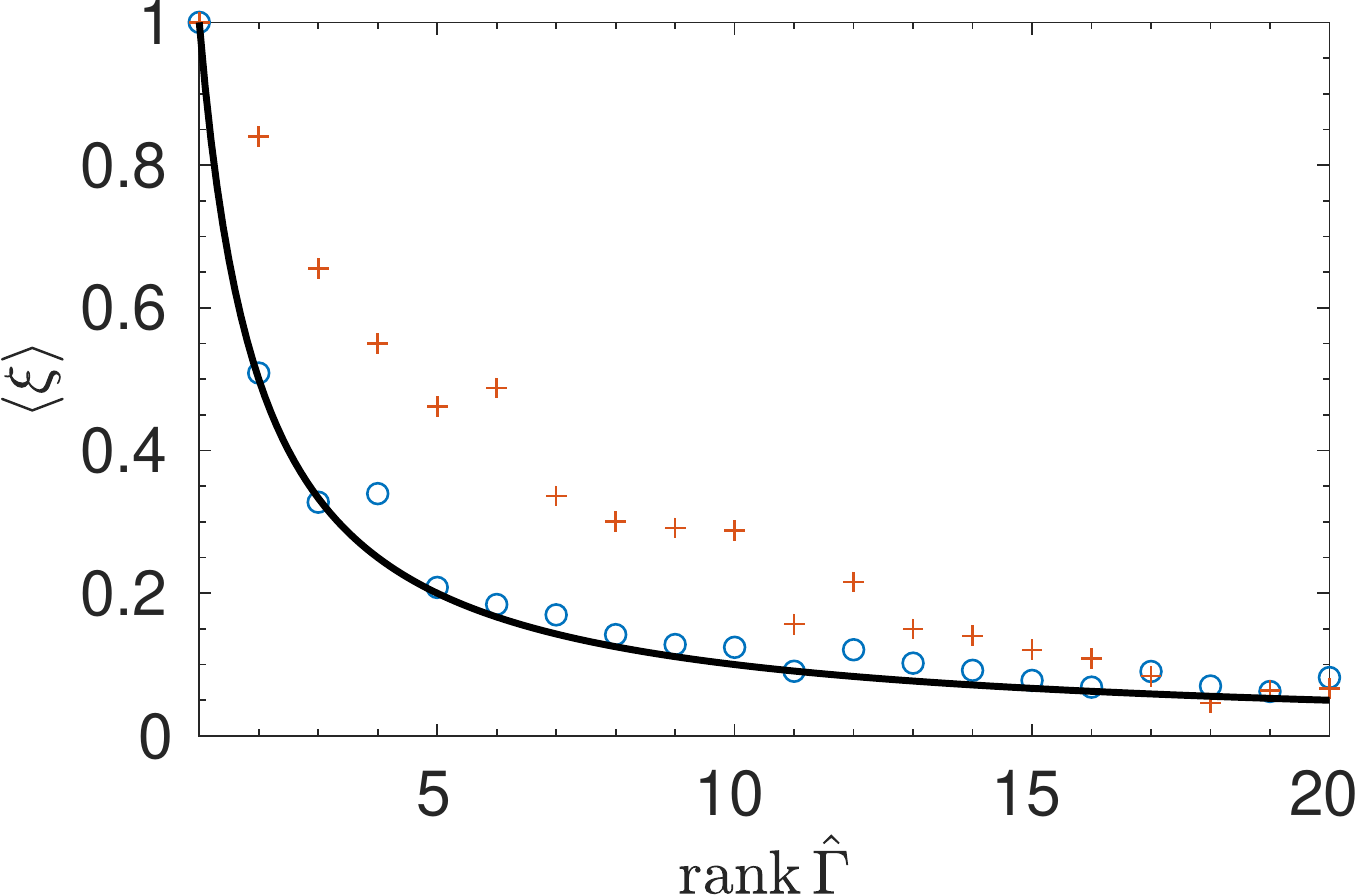}
\caption{Average of the normalized bound $\mean{\xi}$ [dimensionless, see Eqs.~(\ref{eq:LWxi})-(\ref{eq:LWxiav})] as a function of the rank of the decay matrix~$\oHami$. The curve (actually points connected by a curve as a guide to the eye) is the prediction in Eq.~(\ref{eq:xivsrank}) and the circle (plus) symbols mark the numerical results for a random (structured) Hermitian part of the Hamiltonian, see text for details.}
\label{fig:xivsrank}
\end{figure}

Figure~\ref{fig:xivsrank} shows a comparison of the prediction~(\ref{eq:xivsrank}) with numerical results using MATLAB for two different systems. In both cases we construct a $20\times 20$ Hamiltonian matrix in the eigenbasis of the decay matrix~$\oHami$. For the first system (circles) we choose uniformly random numbers $\in [0,1]$ in dimensionless units for the matrix elements of the Hermitian part of the Hamiltonian (describing the closed system dynamics). For the second system (pluses), we use a more structured Hermitian part: a linear tight-binding chain with uniform nearest-neighbor coupling of unity. For the non-Hermitian part (describing the losses) we use in both cases for the first $M$ diagonal elements random numbers picked uniformly from the interval~$(0,2]$.
For the first system Fig.~\ref{fig:xivsrank} demonstrates a very good agreement of the numerical data and the prediction in Eq.~(\ref{eq:xivsrank}). For the second system clear deviation can be observed but still the prediction~(\ref{eq:xivsrank}) qualitatively describes the behavior. Calculations on larger matrices (not shown) confirm this result.

\subsection{Biorthogonal basis}
As nowadays biorthogonal bases are commonly used, we reformulate the Lee-Wolfenstein inequality~(\ref{eq:LWinequality}) in terms of the biorthogonal basis of the Hamiltonian
\begin{equation}\label{eq:biorthogonal}
\oHam\ket{R_\mb} = E_\mb\ket{R_\mb}
\;\;\text{and}\;\;
\bra{L_\mb}\oHam = E_\mb\bra{L_\mb}
\end{equation}
with the right eigenvectors $\ket{R_\mb} = \ket{\psi_\mb}$ and the left eigenvectors $\ket{L_\mb}$. With this definition we can always choose $\braket{L_\ma}{R_\mb} = 0$ if $\ma \neq \mb$ which superficially solves the problem of nonorthogonality. However, it reappears when considering the overlap matrix~\cite{HS17} 
\begin{equation}\label{eq:O}
O_{\ma\mb} = \frac{\braket{R_\mb}{R_\ma}\braket{L_\ma}{L_\mb}}{\braket{R_\mb}{L_\mb}\braket{L_\ma}{R_\ma}} \ .
\end{equation}
Note that in Ref.~\cite{HS17} the transposed matrix is used and that in Refs.~\cite{CM98,FM02} the denominator is missing due to the normalization~$\braket{L_\mb}{R_\mb} = 1$ for all $\mb$. As the choice of normalization is irrelevant for the result of this subsection we use here for convenience as before $\braket{R_\mb}{R_\mb} = 1$ and $\braket{L_\mb}{L_\mb} = 1$ for all~$\mb$ and derive from Eq.~(\ref{eq:O})
\begin{equation}
\frac{|O_{\ma\mb}|}{\sqrt{O_{\ma\ma}O_{\mb\mb}}} = |\braket{R_\ma}{R_\mb}| |\braket{L_\ma}{L_\mb}| \ .
\end{equation}
Using the inequality~(\ref{eq:LWinequality}) for $|\braket{R_\ma}{R_\mb}|$ and also for $|\braket{L_\ma}{L_\mb}|$ (with negative semi-definite decay matrix) we conclude
\begin{equation}\label{eq:LWbi}
\frac{|O_{\ma\mb}|}{\sqrt{O_{\ma\ma}O_{\mb\mb}}} \leq  \frac{\decayrate_\ma\decayrate_\mb}{\Delta_{\ma\mb}^2+(\decayrate_\ma+\decayrate_\mb)^2/4} \ .
\end{equation}
Equation~(\ref{eq:LWbi}) is the reformulation of the Lee-Wolfenstein inequality for a biorthogonal basis. As $1/|\braket{L_\mb}{R_\mb}|$ is a measure of sensitivity of the corresponding eigenvalue under perturbation, see, e.g.,~\cite{GL96}, we can consider Eq.~(\ref{eq:LWbi}) as a bound for certain sensitivity measures. We expect that this might be relevant for response theory~\cite{HS19}.

\section{Wave systems with absorption and radiation}
\label{sec:wave}
This section deals with quantum and electromagnetic systems described not by an effective Hamiltonian but by a wave equation subjected to outgoing-wave conditions.

\subsection{Quantum systems}
\label{sec:QM}
We consider the stationary Schr{\"o}dinger equation for a spinless particle of mass $m$  
\begin{equation}\label{eq:sSG}
\left(-\frac{\hbar^2}{2m}\nabla^2 + V(\vc{x})\right)\psi_\mb(\vc{x}) = E_\mb\psi_\mb(\vc{x}) \ .
\end{equation}
For notational convenience we suppress in the following the dependence on the spatial coordinates~$\vc{x}$. The potential $V$ is assumed to be of finite support (nonzero over a finite region), or at least of short range. We allow the potential $V$ to be complex-valued. For example, for ultracold atoms in optical lattices it is possible to design complex optical potentials~$V$, where $\imagc{V} < 0$ describes an incoherent loss of atoms~\cite{KOA97}. Moreover, losses can occur due to radiation. This is described in $n$ spatial dimensions asymptotically by the outgoing-wave condition
\begin{equation}\label{eq:owc3D}
\psi =\frac{e^{ikr}}{r^\frac{n-1}{2}} \ffA
\end{equation}
with radial coordinate $r$ and wave number $k = \sqrt{2mE}/\hbar$. For bound states, the energy~$E$ is negative and the square root has to be chosen such that $ik < 0$. For quasibound states, the energy is complex and the square root has to be chosen such that the real part of $k$ is positive. In one dimension the amplitude $\ffA$ is a constant, in two dimensions $\ffA = \ffA(\azimuthal)$ with azimuthal angle $\azimuthal$, and in three dimensions $\ffA=\ffA(\polar,\azimuthal)$ with polar angle $\polar$. 

\begin{figure}[ht]
\includegraphics[width=0.65\columnwidth]{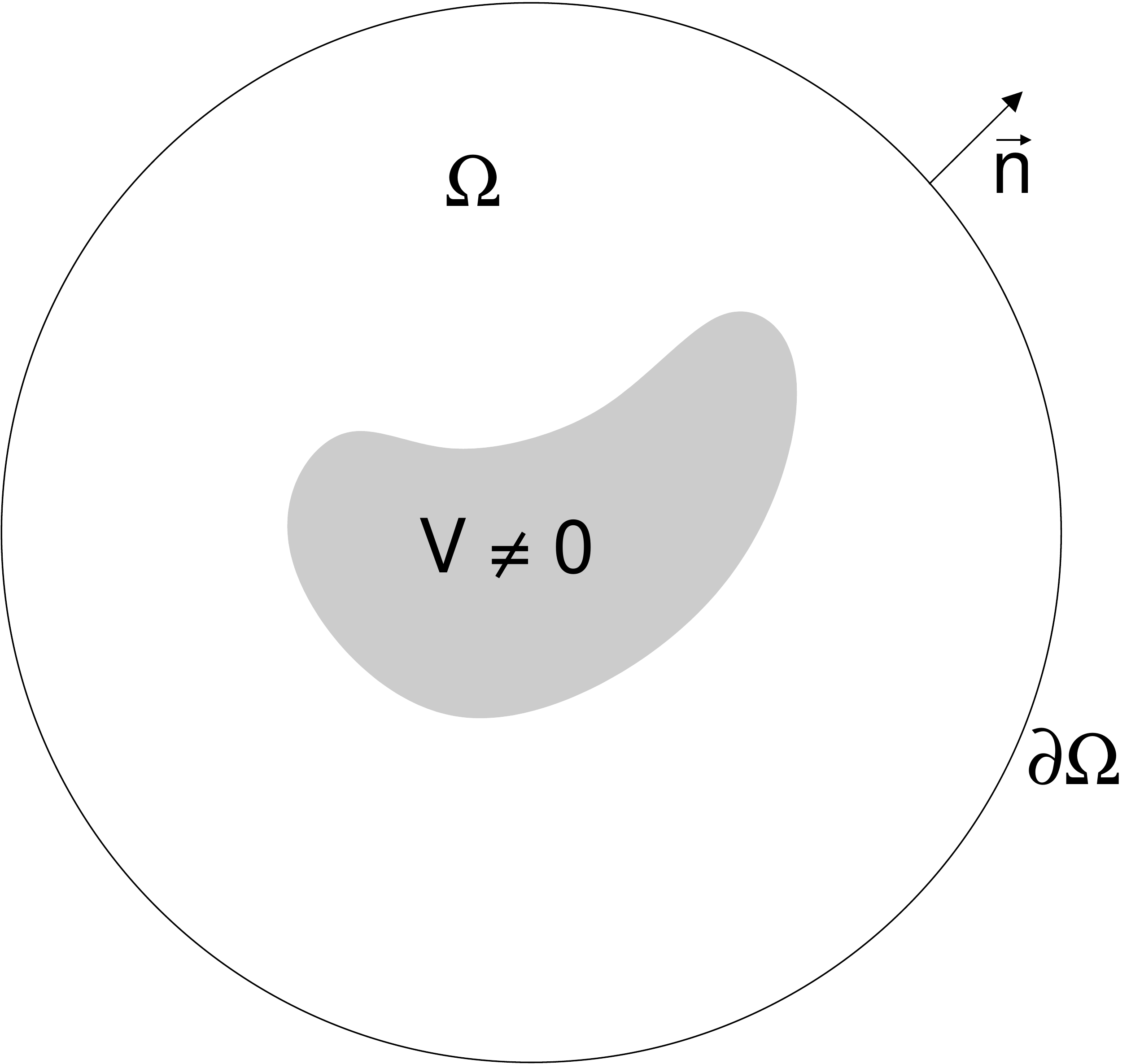}
\caption{Sketch of the region with nonvanishing potential (gray shading) and the volume of integration $\vol$. $\normalv$ is the local normal vector on the surface $\surf$.}
\label{fig:sketch}
\end{figure}

From Eq.~(\ref{eq:sSG}) it is straightforward to obtain
\begin{align}
\nonumber
-i(E_\ma^*-E_\mb)\psi^*_\ma\psi_\mb & = \frac{i\hbar^2}{2m}\nabla( \psi_\mb\nabla\psi^*_\ma-\psi_\ma^*\nabla\psi_\mb)\\
& - 2\,\imag{V}\psi^*_\ma\psi_\mb \ .
\end{align}
Integrating this equation over a finite region~$\vol$, see Fig.~\ref{fig:sketch}, and applying Gauss' theorem we get
\begin{equation}\label{eq:formsQM}
-i(E_\ma^*-E_\mb)\braket{\psi_\ma}{\psi_\mb} = \Hform{\psi_\ma}{\psi_\mb}   
\end{equation}
with the scalar product
\begin{equation}\label{eq:spQM}
\braket{\phi}{\psi} = \int_\vol d^nx\, \phi^*\psi
\end{equation}
and the Hermitian form
\begin{align}
\nonumber
\Hform{\phi}{\psi} & = \hbar\int_{\surf}\dsurf\,\normalv\cdot\frac{i\hbar}{2m}\left(\psi\nabla\phi^*-\phi^*\nabla\psi\right)\\
\label{eq:HfQM}
& - 2\int_\vol d^nx\,\imag{V} \phi^*\psi \ .
\end{align}
$\normalv$ is the local unit normal vector on the surface $\surf$ pointing outwards, see Fig.~\ref{fig:sketch}.

For the special case $\phi = \psi$ we can interpret the Hermitian form in Eq.~(\ref{eq:HfQM}) as follows. The first term on the right-hand side is the flow of the probability current 
\begin{equation}\label{eq:jdef}
\vc{j} = \frac{i\hbar}{2m}\left(\psi\nabla\psi^*-\psi^*\nabla\psi\right)
\end{equation}
through $\surf$ to the exterior region. The second term in Eq.~(\ref{eq:HfQM}) describes the change of probability due to absorption or gain. 

Under what conditions is the Hermitian form~(\ref{eq:HfQM}) positive semi-definite, i.e., $(\psi,\psi) \geq 0$ for all wave functions $\psi$? As the two decay processes of radiation and absorption/gain are independent we have to require both terms in Eq.~(\ref{eq:HfQM}) to be positive semi-definite. 
For the second term this implies $\imagc{V} \leq 0$ for all $\vc{x}$, i.e., there is no gain. If in addition there is also no radiation, i.e., the first term is zero for all states for example for a system with Dirichlet boundary conditions on a surface $\surf$, the Hermitian form~(\ref{eq:HfQM}) is positive semi-definite. From this and the Cauchy-Schwarz inequality~(\ref{eq:Cauchy}) follows directly the Lee-Wolfenstein inequality~(\ref{eq:LWinequality}).

If the first term is nonzero then the reasoning is much more subtle. At first glance it appears to be enough to require that the region~$\vol$ is convex, including the region with $V\neq 0$, and is sufficiently large. The outgoing-wave conditions~(\ref{eq:owc3D}) then give for all energy eigenstates $\psi_\mb$ always a current pointing outward $\vol$, i.e., $\normalv\cdot\vc{j} \geq 0$, and therefore $(\psi_\mb,\psi_\mb) \geq 0$. Unfortunately, from this we cannot conclude that the Hermitian form~(\ref{eq:HfQM}) is positive semi-definite, as there is the possibility that superpositions of energy eigenstates, such as $\psi = \psi_\mb+\alpha\psi_\ma$, exhibit $\normalv\cdot\vc{j} < 0$ at certain positions on the surface~$\surf$. This can indeed happen due to quantum backflow~\cite{Allcock69,Berry10}, the remarkable effect in which a wave packet flows in the direction opposite to its momentum. 
To illustrate this effect, consider the one-dimensional example of a free particle with wave function 
\begin{equation}
\psi(x) = e^{ik_\mb x} + \alpha e^{ik_\ma x} \ .
\end{equation}
A short calculation gives for the corresponding current 
\begin{equation}\label{eq:jex}
j_x = \frac{\hbar}{m}\left\{k_\mb+\alpha^2k_\ma + \alpha(k_\mb+k_\ma)\cos{[(k_\mb-k_\ma)x]}\right\}
\end{equation}
where, for simplicity, we have assumed $k_\ma, k_\mb, \alpha\in\R$. The third term in Eq.~(\ref{eq:jex}) is an interference term originating from the fact that the current~(\ref{eq:jdef}) is a quadratic function of the wave function. Due to the interference the  current can be negative for certain intervals in space even if both wave numbers $k_\ma$ and $k_\mb$ -- and the corresponding momenta $\hbar k_\ma$ and $\hbar k_\mb$ -- are positive. For example, choose $x = \pi/|k_\mb-k_\ma|$, $k_\ma$ small and $\alpha$ large enough.
The quantum backflow has here two related consequences: (i) the probability inside $\Omega$ can temporarily increase, and (ii) the Hermitian form~(\ref{eq:HfQM}) is not positive semi-definite. The nonorthogonality of quasibound states is therefore not constraint by the Lee-Wolfenstein inequality! 

This conclusion seems to be in contradiction with a considerable amount of literature (see~\cite{MRW10} for a review) where open systems, including systems studied in this section~\cite{SSS03}, are modeled with an effective Hamiltonian where the decay operator is deduced to be of the form
\begin{equation}\label{eq:VVT}
\oHami = \couplingm\couplingm^\dagger
\end{equation}
with the $N\times M$ matrix $\couplingm$ and $M = \rank{\oHami}\leq N$. Clearly, $\oHami$ is positive semi-definite is this case, and therefore quantum backflow should be ruled out. The apparent contradiction resolves itself if we take note of the fact that the decay operator in Eq.~(\ref{eq:VVT}) is only locally defined. In general, $\oHami$ in Eq.~(\ref{eq:VVT}) is energy dependent, see also the discussion in Ref.~\cite{SSS03}. Even a weak energy dependence allows for quantum backflow as the interference of a superposition of two eigenstates (eigenstates of formally two different Hamiltonians) is modified which can lead to a temporarily increasing probability.
Another possible pitfall is that the effective Hamiltonian is given in a nonorthogonal basis if considered from the real-space representation. This is, for instance, the case in the perturbation theory in Ref.~\cite{KYW18}. The transformation between this basis and the real-space representation is non-unitary and therefore does not preserve the positive-semi-definiteness of the Hamiltonian.

In the presence of radiation, can a modified version of the Lee-Wolfenstein inequality be derived?
One possibility seems to spatially average over the extent of the region~$\vol$, thereby smearing out the effect of quantum backflow. This is problematic for a number of reasons: (i) This average should be chosen to be state independent. However, Eq.~(\ref{eq:jex}) tells us that even in the simple one-dimensional example the region in which the current is negative depends on the states via $k_\mb-k_\ma$. (ii) The averaging has to be done also for the normalization $\braket{\psi_\mb}{\psi_\mb} = 1$. (iii) The interpretation of the scalar product is no longer obvious. 

In the following we describe a much better way to derive a modified version of the Lee-Wolfenstein inequality. To do so, we choose the region~$\vol$ to be a sphere of radius $R$ containing the region with $V \neq 0$. For one-dimensional systems ($n = 1$) the wave function outside $\vol$ is described by the outgoing-wave condition~(\ref{eq:owc3D}). For $n > 1$ this is true asymptotically for large enough $R$. In this regime,
\begin{equation}
\frac{\partial \psi}{\partial r} = ik\psi \ .
\end{equation}
Plugging this and Eq.~(\ref{eq:owc3D}) into Eq.~(\ref{eq:HfQM}) gives 
\begin{align}
\nonumber
\Hform{\psi_\ma}{\psi_\mb} & = \frac{\hbar^2}{2m}(k_\ma^*+k_\mb) e^{-ik_\ma^*R}e^{ik_\mb R} \braketff{\ffA_\ma}{\ffA_\mb}\\
\label{eq:asymp}
& - 2 \int_\vol d^nx\,\imag{V} \psi_\ma^*\psi_\mb 
\end{align}
with the scalar product
\begin{equation}
\braketff{\ffA}{\ffB} = \int_{\surf} \frac{\dsurf}{R^{n-1}}\, \ffA^*\ffB \ .
\end{equation}
This scalar product does not depend on $R$ and contains only far-field data (hence the subscript \quoting{ff}). 
In the special case $\braketff{\ffA_\ma}{\ffA_\mb} = 0$, the quantity $|\Hform{\psi_\ma}{\psi_\mb}|$ is independent of $R$ because the only contribution in the second term of Eq.~(\ref{eq:asymp}) comes from the region with $V \neq 0$ which is always enclosed by the sphere $\vol$ according to the requirement. In the generic case $\braketff{\ffA_\ma}{\ffA_\mb} \neq 0$ the first term of Eq.~(\ref{eq:asymp}) dominates for quasibound states when $R$ is sufficiently large. 
In this case we get with Eq.~(\ref{eq:formsQM}) 
\begin{equation}\label{eq:ovA}
\braket{\psi_\ma}{\psi_\mb} = \frac{i}{k_\ma^*-k_\mb} e^{-ik_\ma^*R}e^{ik_\mb R}\braketff{\ffA_\ma}{\ffA_\mb}  \ .
\end{equation}
Applying the Cauchy-Schwarz inequality to the scalar product $\braketff{\cdot}{\cdot}$ we obtain
\begin{equation}\label{eq:modLW}
\frac{|\braket{\psi_\ma}{\psi_\mb}|^2}{\braket{\psi_\ma}{\psi_\ma}\braket{\psi_\mb}{\psi_\mb}} \leq  \frac{\decayrate_\ma\decayrate_\mb}{\Delta_{\ma\mb}^2+(\decayrate_\ma+\decayrate_\mb)^2/4}
\end{equation}
with $\Delta_{\ma\mb} = \realc{k_\ma}-\realc{k_\mb}$ and $\decayrate_\mb = -2\,\imagc{k_\mb} \geq 0$. This is very similar to inequality~(\ref{eq:LWinequality}) except for the explicit normalization on the left-hand side to compensate for the $R$-dependence (and therefore for the divergence of the quasibound states in the limit $R\to\infty$) and the appearance of the wave number $k$ instead of the energy $E$. The latter modification makes a difference because of the nonlinear free-space dispersion relation $E = \hbar^2k^2/(2m)$. Interestingly, the right-hand side of inequality~(\ref{eq:modLW}) is larger than the right-hand side of inequality~(\ref{eq:LWinequality}). However, the difference is small in the case of long-lived energy eigenstates with $|\imagc{E_\mb}| \ll |\realc{E_\mb}|$.

For one-dimensional systems $\braketff{\ffA_\ma}{\ffA_\mb} = \ffA_\ma^*\ffA_\mb$ and therefore $|\braketff{\ffA_\ma}{\ffA_\mb}|^2 = \braketff{\ffA_\ma}{\ffA_\ma}\braketff{\ffA_\mb}{\ffA_\mb}$. Hence, the inequality~(\ref{eq:modLW}) is an equality for $n = 1$ similar to the case of rank-1 losses in Sec.~\ref{sec:rank}. Remarkably, this particular case breaks the bound in the original inequality~(\ref{eq:LWinequality}). Note that for one-dimensional systems without absorption there is no need to consider large~$R$ in Eq.~(\ref{eq:modLW}) because the outgoing-wave condition~(\ref{eq:owc3D}) is exact for any $R$.

It should be emphasized that rescuing the Lee-Wolfenstein inequality by its modified version~(\ref{eq:modLW}) comes at a price. The region~$\vol$ is a large sphere and therefore the scalar product is dominated by the behavior of the quasibound states far away from the physical interesting region~$V \neq 0$. Depending on the situation it might be more reasonable to choose the region~$\vol$ close to the region~$V \neq 0$. The Lee-Wolfenstein inequality is then not valid but this can be seen as an advantage if one is interested in strong nonorthogonality and large non-Hermitian effects.

\subsection{Electromagnetic systems}
\label{sec:EM}
Here we consider electromagnetic (quasi-normal) modes as the (damped) time-harmonic solutions of the Maxwell's equations
\begin{equation}\label{eq:Me}
\nabla\times\vc{E} = -\frac{\partial}{\partial t}\vc{B}
\;\;\text{and}\;\;
\nabla\times\vc{H} = \frac{\partial}{\partial t}\vc{D}
\end{equation}
assuming that no free currents are present. We use the constitutive relations
\begin{equation}
\vc{D} = \varepsilon_0\varepsilon\vc{E}
\;\;\text{and}\;\;
\vc{B} = \mu_0\mu\vc{H}
\end{equation}
and restrict ourselves to isotropic and nondispersive media, i.e., $\mu$ and $\varepsilon$ are scalars and frequency independent. We assume that $\varepsilon-1$ and $\mu -1$ are of finite support in three-dimensional space. We allow $\varepsilon$ to be complex-valued in order to describe gain and loss.  
With the identity 
\begin{equation}
\nabla\cdot\left(\vc{a}\times\vc{b}\right) = \vc{b}\cdot(\nabla\times\vc{a}) - \vc{a}\cdot(\nabla\times\vc{b})
\end{equation}
for any vectors $\vc{a}, \vc{b} \in \C^3$ and the Maxwell's equations~(\ref{eq:Me}) we derive
\begin{eqnarray}
\label{eq:em1}
\nabla\cdot\left(\vc{E}_\ma^*\times\vc{H}_\mb\right) & = & \vc{H}_\mb\cdot\left(-\frac{\partial}{\partial t}\vc{B}_\ma^*\right) - \vc{E}_\ma^*\cdot\frac{\partial}{\partial t}\vc{D}_\mb \ ,\\
\label{eq:em2}
\nabla\cdot\left(\vc{E}_\mb\times\vc{H}_\ma^*\right) & = & \vc{H}_\ma^*\cdot\left(-\frac{\partial}{\partial t}\vc{B}_\mb\right) - \vc{E}_\mb\cdot\frac{\partial}{\partial t}\vc{D}_\ma^* \ .
\end{eqnarray}
Adding Eqs.~(\ref{eq:em1}) and~(\ref{eq:em2}) and integrating over the finite region~$\vol$, we arrive at
\begin{equation}\label{eq:formsED}
-i(\omega_\ma^*-\omega_\mb)\braket{\psi_\ma}{\psi_\mb} = \Hform{\psi_\ma}{\psi_\mb}   
\end{equation}
with the complex eigenfrequencies $\omega_\ma$ and $\omega_\mb$ having a positive real part, the six-dimensional vectors
\begin{equation}
\psi = \left(\begin{array} {c}
\vc{E}_\psi\\
\vc{H}_\psi\\
\end{array}\right)
\;,\,
\phi = \left(\begin{array} {c}
\vc{E}_\phi\\
\vc{H}_\phi\\
\end{array}\right) \ ,
\end{equation}
the scalar product
\begin{equation}\label{eq:spEM}
\braket{\phi}{\psi} = \int_\vol d^3x\, \frac{1}{4}\left(\mu_0\mu\vc{H}_\phi^*\cdot\vc{H}_\psi+\varepsilon_0\real{\varepsilon}\vc{E}_\phi^*\cdot\vc{E}_\psi\right) \ , 
\end{equation}
and the Hermitian form
\begin{align}
\nonumber
\Hform{\phi}{\psi} & = \int_{\surf} \dsurf\,\normalv\cdot \frac{1}{4}\left(\vc{E}_\phi^*\times \vc{H}_\psi + \vc{E}_\psi\times \vc{H}_\phi^*\right)\\
\label{eq:HfEM}
& + \int_\vol d^3x\,\frac{1}{4}\varepsilon_0\imag{\varepsilon}i\left(\vc{E}_\phi^*\cdot\frac{\partial\vc{E}_\psi}{\partial t} - \vc{E}_\psi\cdot\frac{\partial\vc{E}_\phi^*}{\partial t}\right) \ .
\end{align}
The interpretation of the scalar product and the Hermitian form is straightforward for the special case $\phi = \psi$ and time-periodic fields with real-valued frequency~$\omega$. The scalar product~(\ref{eq:spEM}) coincides with the total electromagnetic energy in a finite region~$\vol$ averaged over one temporal period $2\pi/\omega$. The Hermitian form~(\ref{eq:HfEM}) reduces to 
\begin{equation}
\Hform{\psi}{\psi} = \int_{\surf} \dsurf\,\normalv\cdot \vc{S} + \int_\vol\frac{\omega}{2}\varepsilon_0\imag{\varepsilon} |\vc{E}|^2
\end{equation}
where 
\begin{equation}
\vc{S} = \frac{1}{2}\real{\vc{E}\times\vc{H}^*}
\end{equation}
coincides with the time-averaged Poynting vector for time-periodic fields. 

For the question of whether the Hermitian form~(\ref{eq:HfEM}) is positive semi-definite we again consider the two processes of radiation and absorption/gain independently. The second term is positive semi-definite if $\imagc{\varepsilon} \geq 0$ for all $\vc{x}$, so again we require that there is no gain. As in the previous section the second term is more subtle. For a sufficiently large and convex region~$\vol$ the outgoing-wave conditions lead to an outward-pointing $\vc{S}$ for individual modes. However, linear combinations allow for optical backflow, the analog of quantum backflow. Optical backflow has been known for a long time, see, e.g., Ref.~\cite{Wolter50}; some modern aspects are discussed in Refs.~\cite{SD10,BBK13}. Analogue to the quantum case in the previous section, the backflow forbids us to conclude that the Hermitian form~(\ref{eq:HfEM}) is positive semi-definite. 
This implies that the matrix $S_{\ma\mb} = \Hform{\psi_\ma}{\psi_\mb}$ is not positive semi-definite (even though $S_{\mb\mb} \geq 0$ for all $\mb$). Note that a similar but not exactly the same matrix (if we ignore dispersion and the regularization of the far-field divergence) has been used for a second quantization scheme of modes in open photonic systems in Ref.~\cite{FHD19}. 

In the presence of absorption only, the Hermitian form~(\ref{eq:HfEM}) is positive semi-definite. From Eq.~(\ref{eq:formsED}) and the Cauchy-Schwarz inequality~(\ref{eq:Cauchy}) follows directly the Lee-Wolfenstein inequality~(\ref{eq:LWinequality}) with 
\begin{equation}\label{eq:DgEM}
\Delta_{\ma\mb} = \realc{\omega_\ma}-\realc{\omega_\mb}
\;\;\text{and}\;\;
\decayrate_\mb = -2\,\imagc{\omega_\mb} \ .
\end{equation}

As in the previous section we ask if in the presence of radiation an alternative nonorthogonality constraint can be found for a large region~$\vol$. Again, $\vol$ shall be a sphere of radius $R$ containing the region with $\varepsilon \neq 1$ and $\mu \neq 1$. For large $R$ we consider the Silver-M{\"u}ller boundary conditions~\cite{AK04}
\begin{equation}
\sqrt{\mu_0}\vc{H}\times\frac{\vc{x}}{|\vc{x}|} = \sqrt{\varepsilon_0}\vc{E}
\;\;\text{and}\;\;
\sqrt{\varepsilon_0}\vc{E}\times\frac{\vc{x}}{|\vc{x}|} = -\sqrt{\mu_0}\vc{H}
\end{equation}
where $\mu = 1 = \varepsilon$ in the far-field region and 
\begin{equation}
\vc{E} = \frac{e^{ikr}}{r}\vc{\zeta}
\end{equation}
with $k = \omega/c$ and $\vc{\zeta}$ depending only on the angles $\polar$ and $\azimuthal$. Inserting these boundary conditions into Eq.~(\ref{eq:HfEM}) gives
\begin{align}
\nonumber
\Hform{\psi_\ma}{\psi_\mb} & = \frac{1}{2}\sqrt{\frac{\varepsilon_0}{\mu_0}}e^{-ik_\ma^*R}e^{ik_\mb R}\braketff{\ffAv_\ma}{\ffAv_\mb} \\
\label{eq:HfEM_asymp}
& + \int_\vol d^3x\,\frac{1}{4}\varepsilon_0\imag{\varepsilon}i\left(\vc{E}_{\ma}^*\cdot\frac{\partial\vc{E}_{\mb}}{\partial t} - \vc{E}_{\mb}\cdot\frac{\partial\vc{E}_{\ma}^*}{\partial t}\right) \ .
\end{align}
with the scalar product
\begin{equation}
\braketff{\ffAv}{\ffBv} = \int_{\surf} \frac{\dsurf}{R^2}\, \ffAv^*\cdot\ffBv \ .
\end{equation}
In the generic case $\braketff{\ffAv_\ma}{\ffAv_\mb} \neq 0$ the first term of Eq.~(\ref{eq:HfEM_asymp}) dominates for quasibound states when $R$ is sufficiently large. In this case we get with Eq.~(\ref{eq:formsED}) 
\begin{equation}\label{eq:ovA}
\braket{\psi_\ma}{\psi_\mb} = \frac{i}{\omega_\ma^*-\omega_\mb}\frac{1}{2}\sqrt{\frac{\varepsilon_0}{\mu_0}}e^{-ik_\ma^*R}e^{ik_\mb R}\braketff{\ffAv_\ma}{\ffAv_\mb}  \ .
\end{equation}
Applying the Cauchy-Schwarz inequality to the scalar product $\braketff{\cdot}{\cdot}$ we get the inequality~(\ref{eq:modLW}) with detuning and decay rates as in Eq.~(\ref{eq:DgEM}). In contrast to the quantum case, we fully recover the Lee-Wolfenstein bound in the limit of a large region~$\vol$ due to the fact that the free-space dispersion relation $\omega = ck$ is here linear. But again the price is that the such obtained scalar product is dominated by the behavior of the modes far away from the physically interesting region with $\varepsilon \neq 1, \mu \neq 1$.

\section{Geometric interpretation}
\label{sec:metric}
One known interpretation of the Lee-Wolfenstein inequality~(\ref{eq:LWinequality}) is that the right-hand side can be understood as spectral overlap of the two states in terms of their associated normalized Breit-Wigner (Lorentz) distributions~\cite{DH70}. 
The geometric interpretation that is presented in this section is very different. We relate each side of the inequality to a distance either in Hilbert space or in complex energy (or frequency or wavenumber) space. 

A metric or distance function $d(x,y)$ between two arbitrary elements $x$ and $y$ of a given set is a mapping to the real numbers with axioms: (i) identity of indiscernibles $d(x,y) = 0$ if and only if $x = y$, (ii) symmetry $d(x,y) = d(y,x)$, and (iii) triangle inequality $d(x,y) \leq d(x,z)+d(z,y)$. From these axioms one can deduce $d(x,y) \geq 0$.

We consider normalized states $\braket{\psi_\mb}{\psi_\mb} = 1$ and rewrite inequality~(\ref{eq:LWinequality}) as
\begin{equation}\label{eq:LWr}
\sqrt{1-|\braket{\psi_\ma}{\psi_\mb}|^2} \geq  \frac{|E_\ma-E_\mb|}{|E_\ma^*-E_\mb|} 
\end{equation}
with the positive branch of the square root and the complex eigenenergies $E_\mb$. The left-hand side is the Hilbert-Schmidt distance of two pure and normalized states $\ket{\phi}, \ket{\psi} \in\Hilbert$ (up to a factor of $\sqrt{2}$, see, e.g.,~\cite{DMM00})
\begin{equation}\label{eq:dHS}
\dHS(\phi,\psi) = \sqrt{1-|\braket{\phi}{\psi}|^2} \ .
\end{equation}
Obviously, this distance is bounded. The maximal value of unity is attained if the two states are orthogonal. The minimal value of zero is attained if the states are collinear. An easy intuitive visualization is provided by vectors in $\R^N$ where $\dHS = |\sin\alpha|$ with $\alpha \in [0,\pi/2]$ being the angle between the two involved vectors.

\begin{figure}[ht]
\includegraphics[width=1.0\columnwidth]{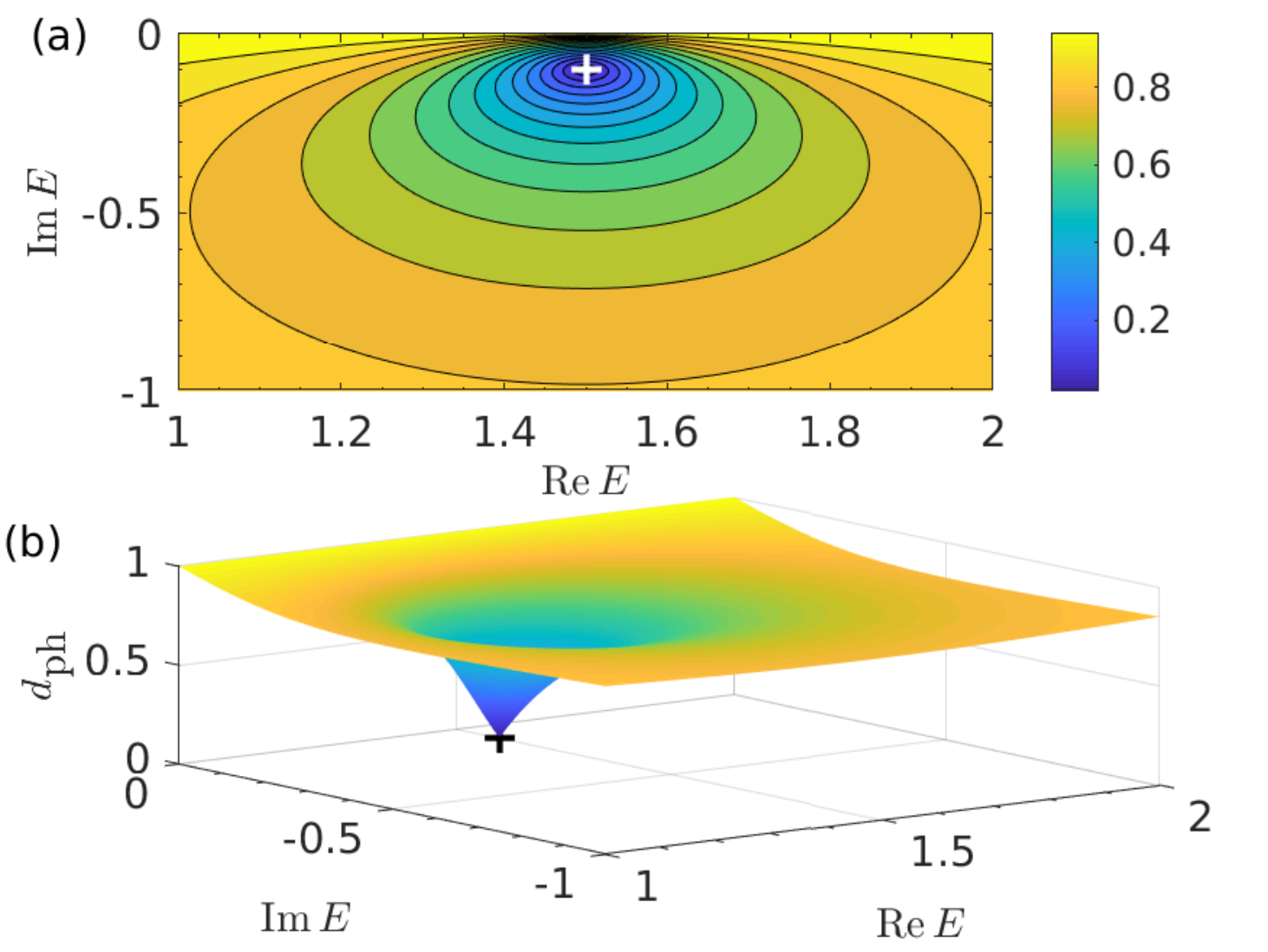}
\caption{Pseudo-hyperbolic distance $\dph(E,E')$ [Eq.~(\ref{eq:dph})] with $E' = 1.5-i0.1$ marked by the plus symbol. (a) Isolines of constant distance and (b) three-dimensional representation. Energies are given in dimensionless units.}
\label{fig:hypdist}
\end{figure}

The right-hand side of Eq.~(\ref{eq:LWr}) can be expressed in two ways. First, we can identify it with the pseudo-hyperbolic distance on the lower half complex plane, see, e.g.,~\cite{Mateljevic18}
\begin{equation}\label{eq:dph}
\dph(E,E') = {\frac{|E-E'|}{|E^*-E'|}} \ .
\end{equation}
By convention, the literature actually considers the upper half plane, see, e.g.,~\cite{DD09}, but the distance is invariant under flipping the imaginary parts of $E$ and $E'$ simultaneously. Hyperbolic distances describe surfaces of constant negative curvature. A distance is called pseudo-distance if the first axiom is relaxed in the sense that still $d(x,x) = 0$ for all $x$, but also $d(x,y) = 0$ is possible for $x \neq y$. This is the case here as in the limit $E, E' \to \infty$ the distance $\dph(E,E') = 0$ even for $E\neq E'$. This is not a problem for our purpose. The same is true for the fact that the distance~(\ref{eq:dph}) is not additive along geodesics~\cite{Mateljevic18}. Also the distance $\dph(E,E')$ is bounded and lies in the interval $[0,1]$. An illustration is provided in Fig.~\ref{fig:hypdist}.

With Eqs.~(\ref{eq:dHS}) and (\ref{eq:dph}) the Lee-Wolfenstein inequality~(\ref{eq:LWinequality}) can be written as a relation between distances in Hilbert space and complex energy space
\begin{equation}\label{eq:LWD1}
\dHS(\psi_\ma,\psi_\mb) \geq \dph(E_\ma,E_\mb) \ .
\end{equation}
Hence, states that are \quoting{spectrally close} -- now rigorously defined by the distance function~$\dph$ -- can be also close in Hilbert space, i.e., strongly nonorthogonal.
In the context of Fig.~\ref{fig:hypdist}(b), this inequality means that, for fixed $E' = E_\mb$ and variable $E = E_\ma$, the distance $\dHS$ can take on values that are located on or above the surface $\dph(E,E')$.

Alternatively, we can consider the hyperbolic distance on the lower half plane (the Poincar\'e metric, see, e.g.,~\cite{DD09}) 
\begin{equation}\label{eq:dP}
\dP(E,E') = 2\text{arctanh}{\frac{|E-E'|}{|E^*-E'|}} \ .
\end{equation}
This distance function is unbounded and additive along geodesics. For small distances, the pseudo-hyperbolic distance and the hyperbolic distance coincide, up to the obvious factor of 2. With Eqs.~(\ref{eq:dHS}) and (\ref{eq:dP}) the Lee-Wolfenstein inequality~(\ref{eq:LWinequality}) can be written as 
\begin{equation}\label{eq:LWD2}
\dHS(\psi_\ma,\psi_\mb) \geq \tanh{\left[\frac{\dP(E_\ma,E_\mb)}{2}\right]} \ .
\end{equation}
For the single-decay channel case the inequalities~(\ref{eq:LWD1}) and~(\ref{eq:LWD2}) turn into equalities for all energy eigenstates.

The formulation of the nonorthogonality bound in terms of distance functions in Eqs.~(\ref{eq:LWD1}) and~(\ref{eq:LWD2}) is relevant for two reasons: (i) It gives the interpretation \quoting{spectrally close} a precise meaning. (ii) It can be extended to $n > 2$ energy eigenstates, e.g., by
\begin{equation}\label{eq:LWD1e}
\sum_{\mb=1}^{n-1}\dHS(\psi_\mb,\psi_{\mb+1}) \geq \sum_{\mb=1}^{n-1}\dph(E_\mb,E_{\mb+1}) \ .
\end{equation}
This inequality compares the lengths of two polygon courses, one in Hilbert space and one in complex energy space. An expression similar to the left-hand side of Eq.~(\ref{eq:LWD1e}) has been used in Ref.~\cite{KYW18} as a cost function to search for (higher-order) exceptional points in parameter space. 

\section{Discussion and conclusions}
\label{sec:discussion}
Various aspects of nonorthogonality constraints in open quantum and wave systems have been discussed. 
Starting with an introduction of Hermitian forms we have shown that positive semi-definiteness of these forms is sufficient to derive the Lee-Wolfenstein bound for the overlap of two energy eigenstates. 
We have revealed that for the case of single-channel decay the nonorthogonality is maximal in the sense that the Lee-Wolfenstein inequality turns into an equality. For increasing rank of the decay operator the average nonorthogonality decreases.  
We have also shown how the Lee-Wolfenstein inequality can be reformulated for the case of a biorthogonal basis. The reformulated version can be considered a bound for sensitivity measures.

We have demonstrated that the condition of positive semi-definiteness does not apply to quantum and wave systems with radiation. The reason is the phenomenon of quantum or optical backflow. The lack of an upper bound for the overlap allows larger effects of nonorthogonality.

The fact that the backflow spoils the positive semi-definiteness of the Hermitian form or the decay operator is itself an interesting finding that is also relevant for the recent experiments on sensors based on exceptional points~\cite{COZ17,HHW17,DLY19}. It had been shown that the positive semi-definiteness implies a restriction for the operation of passive devices indicating that gain is needed for practical applications~\cite{Wiersig14b,Wiersig16}. Our finding implies that there is a chance that the usage of gain might be avoided.  

We have shown that the Lee-Wolfenstein inequality can be rescued by extending the range of integration for the scalar product to the far-field region. Moreover, for quantum systems one has to replace energies by wave numbers in the inequality. For the case of three-dimensional electromagnetic systems the corresponding replacement of frequencies by wave numbers is irrelevant because of the linear free-space dispersion relation. 

Finally, we have introduced a geometric interpretation of the nonorthogonality bound. We have related the Lee-Wolfenstein inequality to a relation of the Hilbert-Schmidt distance, measuring the nonorthogonality, to the (pseudo-)hyperbolic  distance on the lower half plane of complex energies, measuring the spectral closeness.

\acknowledgments 
Fruitful discussions with H. Schomerus and P.~T. Kristensen are acknowledged.


%

\end{document}